\documentclass[print]{revtex4}
\usepackage{amsmath,amssymb}
\usepackage{graphicx}

\draft
\begin{document}

\title{Classical and quantum depinning of a domain wall with a fast varying spin-polarized current}

\author{Xin Liu$^{a,b}$\footnote{Electronic address:liuxin03@mail.nankai.edu.cn}, Xiong-Jun
Liu$^{c}$ and Zheng-Xin Liu$^{a,b}$} \affiliation{a. Theoretical
Physics Division, Nankai Institute of Mathematics, Nankai
University, Tianjin 300071, P.R.China\\
b. Liuhui Center for Applied Mathematics, Nankai University and
Tianjin University, Tianjin 300071, P.R.China\\
c. Department of Physics, National University of Singapore, 2
Science Drive 3, Singapore 117542}

\begin{abstract}
We study in detail the classical and quantum depinning of a domain
wall (DW) induced by a fast-varying spin-polarized current. By
confirming the adiabatic condition for calculating the spin-torque
in fast-varying current case, we show that the time-dependent spin
current has two critical values that determine the classical
depinning of DW. This discovery successfully explains the recent
experiments. Furthermore, a feasible way is proposed to lower the
threshold of spin currents and control the direction of DW¡¯s
motion. Finally, the quantum properties for the depinning of DW are
also investigated in this paper.
\newline

\end{abstract}

\date{\today}
\maketitle

\subsection*{I. INTRODUCTION} \indent
Recently, considerable attention has been paid to displace a domain
wall (DW) by a spin-polarized current in ferromagnetic nanowire and
thin film. Upon the view of basic physics, this phenomenon is
derived from the interaction between the conducting electrons and
the local moment of the DW, which exerts a spin-torque on the DW in
the adiabatic condition. By now the adiabatic spin torque is
calculated in the constant current case so that the previous
theoretical works are mostly focused on the relation between the
constant current and the motion of the DW \cite{modify, S Z, sd,
uniform, another}. Very recently, however, several experiments
discovered many intriguing phenomena by using a fast-varying
spin-polarized current \cite{ac, sub}, which motivates us to
theoretically study the dynamics of DW in such a fast-varying
current case.

In this paper, we first confirm the validity of adiabatic
spin-torque in the case of using a fast-varying spin-polarized
current, and then study the dynamics of classical and quantum
depinning of DW in the biaxial ferromagnet. The validity of
adiabatic condition can be verified in the fast-varying current
case by the method proposed in Ref. \cite{modify}, thus the
adiabatic spin-torque can be readily obtained. Based on this
result we shall show the fast-varying characteristic of the
current is one of the key points to determine the classical
depinning of DW, even if the time-varying of the current only
persists in a short time, $10^{-10}\backsim 10^{-9}s$. Our
discovery also provides a self-contained theory for the insights
proposed by L. Berger \cite{berger} a few years ago. The quantum
depinning of the DW is also studied in this paper.

The paper is organized as follow. In Sec. II, we confirm the
adiabatic condition in the case of fast varying (increasing or
falling) current and ac current, and prove that the spin-torque
obtained in Ref. \cite{modify} is also approved in our case. We then
calculate in detail the system's Lagrangian that determines the
motion of DW. In doing this, we use the solution obtained in Ref.
\cite{last} instead of the traditional trial function \cite{walk} of
DW to completely describe the interaction Hamiltonian, which is
essential to study the quantum depinning of DW. In Sec. III, with
the Lagrangian obtained in Sec. II, we study the classical depinning
of the DW. The results obtained here are good in agrement with the
recent experiments. The quantum depinning of the DW is investigated
in Sec. IV. Finally, we briefly conclude our results in Sec. V.

\subsection*{II MODEL} \indent
First, let us begin with the adiabatic condition in the case of
the fast-varying spin-polarized current which propagates in a
one-dimensional biaxial ferromagnetic material. We assume that the
average velocity of all ballistic electrons is $\bar{v}$ so that
$ne\bar{v}=J_e$, where $n, e, J_e$ are the density of the
conducting electrons, charge of an electron and charge current
density, respectively. The effective dynamical equation of the
electron spin can be written as follow \cite{modify}
\begin{eqnarray}\label{spin}
i\hbar \frac{\partial \psi}{\partial t}=-J\mathbf{M}(x(t))\cdot
\mathbf{\sigma}\psi,
\end{eqnarray}
where the spin-wave function $\psi=(\phi_1, \phi_2)^T$, $J$ is the
Hunds rule coupling or in general of the local exchange and
$\mathbf{M}$ is the local magnetization. The $2\times2$ matrix
$\mathbf{M}\cdot \mathbf{\sigma}$ can be diagonalized with a local
spin rotation $\phi_{\alpha}=U_{\alpha \beta}\psi_{\beta}$. Thus the
above equation yields
\begin{eqnarray}\label{spintime}
i\hbar \frac{\partial}{\partial
t} {\left(
\begin{array}{lc}
\phi_{1}\\
\phi_{2}
\end{array}\right)}={\left(
\begin{array}{lc}
-JM_{s} & 0\\
0 & JM_{s}\end{array}\right)}{\left(
\begin{array}{lc}
\phi_{1}\\
\phi_{2}
\end{array}\right)}+H^{'}_{\alpha \beta}\phi_{\beta},
\end{eqnarray}
where $H'=-i\hbar U^{+}\partial_{t}U$ is given by
\begin{eqnarray}\label{H}
H'=\frac{-i\hbar \bar{v}}{2} {\left(
\begin{array}{lc} \-i\cos\theta
\partial_{x}\phi &
-\partial_{x}\theta+i\sin\theta\partial_{x}\phi\\
\partial_{x}\theta+i\sin\theta\partial_{x}\phi & i\cos\theta
\partial_{x}\phi \end{array}\right)},
\end{eqnarray}
and can be studied with perturbation theory. The transition rate
$\Gamma$ between the two spin states can be calculated in the
interaction picture \cite{perturbation}
\begin{eqnarray}\label{rate}
\Gamma&=&|\int_{0}^{t} H'_{I12} dt|^2\nonumber\\
&=&|\sum_{k=0}^{\infty}\frac{-i\exp(-i2JMt)}{2JM}[\frac{-i\hbar}{2JM}\partial_{t^{'}}]^k
H_{12}^{'}|_{0}^{t}|^2,
\end{eqnarray}
where $H'_I$ is the perturbation Hamiltonian $H'$ in interaction
picture. To verify the adiabatic condition, we can compare the first
and second terms in above equation. Considering a fast linearly
rising current $J_e=ne\bar{v}=neat$ in our case, where $a$ is the
average acceleration of all conducting electrons, we find
\begin{eqnarray}\label{compare}
\frac{|\Gamma_{k=1}|}{|\Gamma_{k=0}|}=\frac{a\hbar}{2\bar{v}JM}.
\end{eqnarray}
Obviously the transition rate $\Gamma$ mostly depends on the term
$k=0$ unless the term $k=1$ is not much smaller than the one $k=0$,
which case, however, requires $\frac{dJ_e}{dt}\sim J_e/\hbar$
according to Eq. (\ref{compare}). In the recent experiments
\cite{sub, first}, because the maximal current density is about or
smaller than $10^{12}(A/m^2)$ and generally
$\frac{dJ_e}{dt}\leq10^{24}(A/sm^{2})\ll J_e/\hbar$, the adiabatic
condition is satisfied. For the ac current case, we note that the
adiabatic condition can also be satisfied unless the ac current's
frequency $\omega_{c}$ is large enough that $\hbar \omega_{c}
\simeq2JM$. As in experiment \cite{ac}, $\omega_c$ is equal to the
eigenfrequency $\omega$ of the pinning potential, which is
independent of the energy difference between spin-up and spin-down.
Generally when $\omega_c \simeq \omega$, the value $\hbar\omega_c$
is far different from $2JM$, thus the adiabatic condition is also
approved in the ac current case. As a result, we can safely extend
the adiabatic spin-torque to the fast-varying current case. At the
same time, according to above discussion, the magnetoresistance will
have two peaks with the frequency's variety. One corresponds to the
resonance of pinning potential and the other corresponds to the
resonance of transition.

Now we proceed to study the dynamics of DW interacted with a
fast-varying spin-polarized current. We consider an elongated
ferromagnetic sample (or a microfabricated magnetic wire) which
has an effective-one-dimensional (1D) DW structure. When a
spin-polarized current propagates in this sample, together with an
external field $H_{ext}$ along the $x$ axis, the energy of this
system reads \cite{modify, uniform}
\begin{eqnarray}\label{ham}
E&=&\mathcal{A}\int d^{3}x (E_m+E_{int}+E_{ext}),\nonumber \\
E_m&=&2A((\nabla\theta)^{2}+\sin\theta(\nabla\phi)^{2})-\nonumber\\
&&-HM_s(\sin\theta \cos\phi)^{2}+H_{\bot}M_s\cos^{2}\theta, \nonumber \\
E_{int}&=&\frac{\hbar}{2e} J_e\nabla\phi(1-\cos\theta), \ \ \ \
E_{ext}=\vec{H}_{ext}\cdot\vec{M},
\end{eqnarray}
where $\mathcal{A}$ is the cross section of the sample, $M_{s}$ is
the saturation magnetization, $M_{z}=M_{s}\cos \theta$,
$M_{x}=M_{s}\sin \theta \cos \phi$, $H$ and $H_{\bot}$ are
anisotropy fields which respectively correspond to easy $x$ axis
and hard $z$ axis, $e$ is the electronic charge, $E_{int}$ is the
interaction Hamiltonian between the magnetization and the current
on the adiabatic assumption \cite{modify} which is the key point
to understand the dynamics of the magnetic interacting with the
spin-polarized current. Although the interaction Hamiltonian
$E_{int}$ is obtained in the constant current system, it is
applicable for the fast-varying current case, since the adiabatic
condition can still satisfied in present case.


Before calculating the Lagrangian of the system, we shall go over
the traditional soliton solution called trial function introduced
by Walker \cite{walk}, which will be helpful to understand our
later discussion on the DW's depinning. The trial function is
\begin{eqnarray}\label{trial}
\phi=\phi(t), \ln\tan\frac{\theta}{2}=c(t)(x-\int^{t}_{0} v(\tau)d
\tau).
\end{eqnarray}
\begin{figure}[htbp]
\includegraphics[width=0.7\columnwidth]{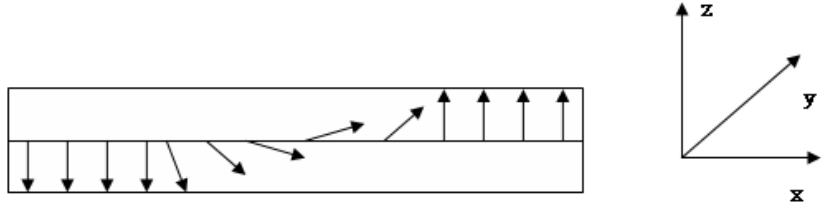}
\caption{A configuration of magnetization described by trial
solution. The magnetization is almost in $x-z$ plane.}\label{}
\end{figure}
Based on above formula the $E_{int}$ is always zero, because
$\phi$ is independent on space coordinate. However, since it is
not the Lagrangian but the variation of Lagrangian to decide the
dynamical equation of the DW, the above solution can safely
describe the classical motion of the DW with the spin-torque
considered \cite{S Z}. But for quantum tunneling phenomenon, the
tunneling rate is decided by the system's Lagrangian, so the
space-dependent character of the solution $\phi$ will be
important. In trial function solution, one should use the sphere
coordinate, which has an artificial singularity at south pole or
north pole, say, in such case we cannot find a complete coordinate
system to include both south and north pole. Noting that the
spin-torque is a topological term derived from the gauge field in
spin representation \cite{modify}, in order to cancel the
singularity, one can take the coordinate neighbour-hoods, i.e. two
sets of coordinates, to describe the topological character of the
system \cite{yang}. In the trial function, the magnetization
points to the north pole when $x\rightarrow -\infty$ and to the
south pole when $x\rightarrow \infty$, see Fig. 1. This in fact
connects the north and south poles in the same position in the
sphere coordinate. Under the description of the incomplete
coordinate system, the solution of trial function always leads to
a zero topological term $E_{int}$. Therefore, to calculate the
Lagrangian of our system, we consider the solution obtained in
\cite{last}
\begin{eqnarray}\label{own}
&&\sin\theta (\frac{\partial \phi}{\partial t}+b_{J} \frac{\partial
\phi}{\partial x})=-\gamma \cos\theta H_{\bot}, \nonumber \\
&&\phi=2\arctan[\exp(\frac{x-q(t)}{\Delta(t)})],
\end{eqnarray}
\begin{figure}[htbp]
\includegraphics[width=0.7\columnwidth]{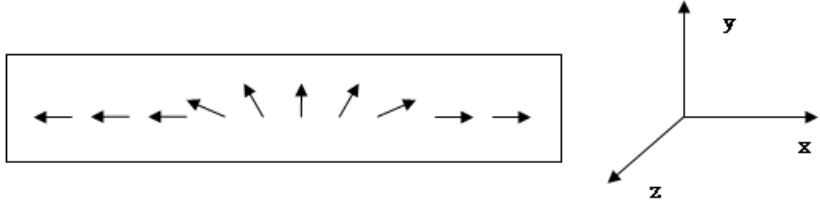}
\caption{A configuration of magnetization described by trial
solution. The magnetization is almost in $x-y$ plane.}\label{}
\end{figure}where $\gamma$ is the gyromagetic ratio, the coefficient of the
spin-torque $b_J=\frac{\mu_{B}PJ_e}{eM_{s}}$ with $P$ and $\mu_B$
are spin polarization of the current and Bohr magneton,
respectively. $q(t)$ is the center of the DW as a kink soliton and
$\Delta(t)$ is the width of the DW. In this solution, the
magnetization lies closely in the $x-y$ plane, so the singularity
in south or north pole is readily cancelled (Fig. 2). Putting this
solution (\ref{own}) into the DW's Lagrangian (\ref{ham}) and
integrating over it, we get the form of the Lagrangian in present
of spin-torque
\begin{eqnarray}\label{lag}
L&=&S^0-\int E=\frac{1}{2}m(\dot{q}+b_{J})^2+Q\varphi-(2A\Delta^{-1}+2K\Delta)\nonumber\\
&&=\frac{1}{2m}(P+QA)^2+Q\varphi-(2A\Delta^{-1}+2K\Delta)\nonumber\\,
&&S^0=\mathcal{A}\int dx\frac{\hbar S}{a^3}\dot{\phi}(\cos\theta-1),
\end{eqnarray}
where $m=(2\pi\gamma^{2}\Delta)^{-1}$ is the effective mass of a
DW, $Q$ is the effective magnetic charge at the center of the DW,
and $\nabla\varphi=H_{ext}$, $A=mb_J/Q$.  From the Eq.
(\ref{lag}), it is clear that the effect of the spin-polarized
current on a DW can be understood as a charged particle interacts
with external magnetic field described by the vector potential
$A$. Since the width $\Delta$ of the DW only changes slightly
during the motion of the DW, we shall assume that $\Delta$ keeps
constant and can be neglected from DW's lagrangian in the
following discussion.

\subsection*{III CLASSICAL DEPINNING} \indent

With the above discussion and considering the damping term
$\frac{\alpha}{M_{s}} \mathbf{M} \times \frac{\partial
\mathbf{M}}{\partial t}$ in present system, the Euler equation of
the Lagrangian (\ref{lag}) is obtained
\begin{eqnarray}\label{euler1}
m\ddot{q}+Q\dot{A}+\alpha_{0} \dot{q}=QH_{ext}.
\end{eqnarray}
Noting that vector potential $A$ is proportional to the current
$J_e$, Eq. (\ref{euler1}) clearly shows that it is the $variation$
of the adiabatic spin-torque that exerts a force to the DW.
Therefore, with a damping term considered, if the applied current
is constant, it cannot maintain the motion of the DW without
external magnetic field. This result coincides with that in Ref.
\cite{S Z} where the terminal velocity is found to be independent
of the spin torque for the constant-current case.

So far our discussion is based on an ideal sample. Practically,
the defects in samples can pin the domain wall by a pinning
potential. In the following, we study the classical depinning of
the DW induced by a rising dc current, with the defects
considered. For convenience, the pinning potential is set by
\begin{eqnarray}\label{clpin}
V(x)=\frac{1}{2}m\omega^2 x^2\theta(\xi-|x|),
\end{eqnarray}
where $\theta(\xi-|x|)$ is the step function and $\xi$ is the half
width of the pinning potential. First, we consider applying a
rising dc current on the sample without external magnetic field.
For example, the current is assumed to linearly rise with time,
$A=Kt (K>0)$. Then the dynamical equation of the DW reads
\begin{eqnarray}\label{euler2}
m\ddot{q}+\alpha_{0} \dot{q}+Qk+m\omega q\theta(\xi-|q|)=0,
\end{eqnarray}
with $\alpha_0=2M_s\alpha /\gamma \Delta$. For overdamping
($\frac{\alpha_0}{2m}=\frac{\alpha M_s \gamma}{\pi} \gg\omega$),
the solution of the Eq. (\ref{euler2}) is
\begin{eqnarray}\label{sol1}
q(t)=C(1-\exp(-\frac{\omega^2}{\alpha_0}t)),
\end{eqnarray}
where $C=\frac{QK}{m\omega^2}$ is the maximum displacement of the DW
and proportional to $\frac{dj_e}{dt}$. To depin the DW, $C$ must be
larger than $\xi$ and the threshold of depinning time $t_{0}$
satisfies $\xi=C(1-\exp(-\frac{\omega^2}{\alpha_0}t_{0}))$. Thus,
the minimum depinning current $J_{min}$ for different $C$ is
\begin{eqnarray}\label{jcr}
&&J_{min}(C)=\int_{0}^{t_{cr}} d(J_e)
=-\frac{CeM_s\alpha_0}{\mu_{B}P}\ln(1-\frac{\xi}{C}),\nonumber\\
&&J_{min}^{cr}=J_{min}(\infty)=\frac{2\alpha
eM_s^2\xi}{\gamma\Delta\mu_B P}.
\end{eqnarray}
\begin{figure}[htbp]
\includegraphics[width=0.8\columnwidth]{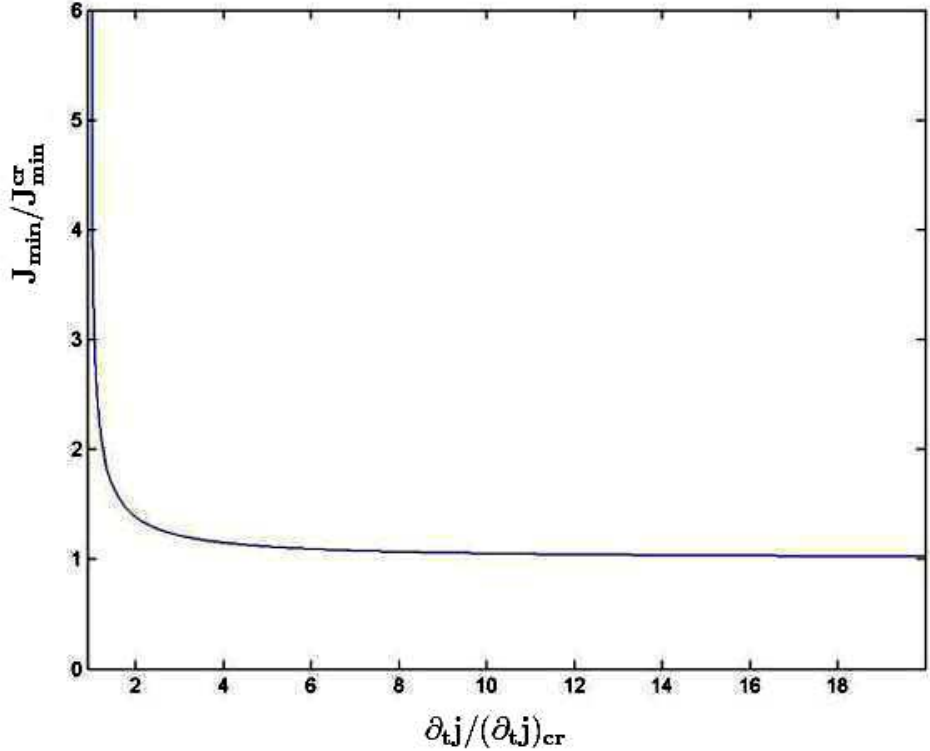}
\caption{Abscissa $\partial_{t}j/(\partial_{t}
j)_{cr}=\frac{dj_e}{dt}/(\frac{dj_e}{dt})_{cr}$. This figure shows
that $J_{min}(C)$ is fast close to $J_{min}^{cr}$ with the rising
value of $C$}\label{}
\end{figure}
As $C \ln(1-\frac{\xi}{C})=-\xi$ when $C=\infty$, it is easily to
see that $J_{min}(C)$ will be fast close to $J_{min}^{cr}$ with
the rising value of $C$ (Fig. 1). Accordingly, a current above
$J_{min}^{cr}$ is necessary for depinning. From above discussion
we conclude that there are two critical values to determine the
depinning of DW by spin-polarized current. The first is $(\frac{d
J_e}{d t})_{cr}=\frac{eM_s\xi \omega^2}{\mu_B P}(\sim C_{cr})$
which assures that the force exerted on DW is big enough for
depinning. The other is $J_{min}^{cr}=\frac{2\alpha
eM_s^2\xi}{\gamma\Delta\mu_B P}$ which indicates the interaction
time should be long enough to be dipin the DW. The two critical
values are solely dependent on the properties of the ferromagnet.
Our results are good in agrement with the recent experiment
\cite{sub}. In this experiment, they observed an important
phenomenon that the critical current $J_{cr}$ is independent of
the current pulse durations. This can be explained by our result
that $J_{cr}$ is determined by the properties of the ferromagnet.
To change the current pulse duration in the experiment is
equivalent to change $\frac{d J_e}{dt}$, e.g. if one use a small
$\frac{d J_e}{dt}$ (but still larger than $(\frac{d J_e}{d
t})_{cr}$), a large pulse duration (or $J_e$) will be required to
depin the DW. We should also emphasize that, except for the width
of the pinning filed $\xi$, the parameters in the equation
$J_{min}^{cr}=\frac{2\alpha eM_s^2\xi}{\gamma\Delta\mu_B P}$ are
observable. As a result, this equation provides us an feasible way
to measure the width of the pinning potential in the ferromagnet.

Subsequently, we study the scheme to lower the threshold $J_e$ of
spin currents by applying both current and external filed
simultaneously. Eq. (\ref{lag}) shows that $H_{ext}=-\nabla \phi$
has the same effect with $\partial_{t} A$. The dynamical equation
of the DW in this case is obtained as
\begin{eqnarray}\label{euler3}
m\ddot{q}+\alpha_{0} \dot{q}+Q(k+H_{ext})+m\omega
q\theta(\xi-|q|)=0.
\end{eqnarray}
According to Eq. (\ref{euler3}),  by applying an external field
lower than critical reverse field $H_{cr}$ as well as a fast-rising
current opposite to the external field, we can obtain an effective
way to depinning the DW with smaller current without lack of the
precision of the control. For example, when the external magnetic
field is set as $H_{ext}=\frac{9}{10} H_{cr}$, the threshold current
will be reduced by factor a 10. This may provide us an effective way
to implement a more dense magnetic memory, since we know that the
current can control the DW in a smaller spatial region than magnetic
field can. For a more dense ferromagnetic memory disk than
present-day's, when the magnetic field applied on the disk is
smaller than $H_{cr}$, it cannot change the information in the disk
by itself but reduce the current threshold. The information is still
controlled by the current, but in a easier way. Nevertheless, if the
fast-rising current is in the same direction with the applied
magnetic field, the threshold of the current will be larger, and the
motion of DW is harder to induce. Our conclusion can also explain
another important phenomenon observed in the experiment \cite{sub}:
when the applied external field is smaller than $H_{cr}$, the
observed direction of DW displacement will be dependent on the
external field. We shall give a more detailed discussion in the
following. We consider the case that the current is in the same
direction with the external magnetic field. Furthermore, we note
$\dot{A}=K_{+},\ \ K_{+}>0$ to describe the rising process of the
spin current, and note $\dot{A}=-K_{-},\ \ K_{-}>0$ to describe the
falling process. According to the Eq. (\ref{euler2}), the total
effective field $H_{eff}$ reads
\begin{eqnarray}\label{heff}
H_{eff}=H_{ext}\mp K_{\pm},
\end{eqnarray}
which means that although the current is in the same direction
with $H_{ext}$ during the process, the rising/falling currents try
to induce the displacement of DW in the opposite/same direction
with that induced by the external magnetic field. Thus, rising and
falling currents separately counteract and enhance the DW's
depinning. Particularly, we may assume that $H_{ext}=\frac{1}{2}
H_{cr}$ and obtain
\begin{eqnarray}\label{rf}
J_{cr+}=\frac{3}{2}J_{cr} \ \ J_{cr-}=\frac{1}{2}J_{cr},
\end{eqnarray}
where $J_{cr}$ is the threshold current without external field,
$J_{cr+}$ and $J_{cr-}$ correspond to the threshold currents for
the rising-current case and falling-current case, respectively. In
experiment \cite{sub}, because the charge current is near
$J_{cr}$, when the external magnetic field is applied, rising
current cannot depin the DW but falling current can. This is why
in experiment the direction of DW displacement is solely dependent
on the external magnetic field.

The effective magnetic field generated by the fast rising current
can be readily estimated by using the materials parameters of
permalloy: $H_{\bot}=4\pi M_{s}=1.0\times 10^{4}$Oe
(demagnetization field has the same effect with the hard
anisotropy field as we have discussed), $H=10$Oe, $A=1.75 \times
10^{-11}J/m,$ $\gamma=1.75\times 10^{7}$(Oe)$^{-1}s^{-1}$, $P=0.5$
\cite{S Z}, rising time is $0.32ns$, $J_{cr}=6.7\times 10^6A/cm^2$
\cite{sub}. The effective magnetic filed of this rising current is
then obtained $H_{eff}=2.36$Oe.

Recently, some authors concentrated on the ac current applying on
the DW \cite{ac, acth}. They concluded that the threshold current
in ac case is smaller than dc case. However it is difficult to
control the direction of the DW's motion because the pinning
potential is symmetry without external magnetic filed in ac case.
According to our result this difficulty can be resolved by
applying an external magnetic field $H_{ext}$ lower than $H_{cr}$.
This result can also be explained that the external magnetic field
breaks the symmetry of the pinning potential and then controls the
direction of DW's displacement. Let's assume the field
$H_{ext}=\frac{1}{2} H_{cr}$ along $x$ axis. The modified
potential then reads
\begin{eqnarray}\label{ac}
V=\frac{1}{2}m\omega^2(x-\frac{1}{2}\xi)^2\theta(\xi-|x|).
\end{eqnarray}
Depinning occurs at the amplitude of displacement $X\geq\xi/2$ and
the threshold current is $J^{ac}_{cr}/2$ ($J^{ac}_{cr}$ is the
threshold without external field). The displacement of the DW must
be in the same direction with the external field if only the
current $J_e$ satisfies $\frac{J^{ac}_{cr}}{2}<J_e\leq \frac{3
J^{ac}_{cr}}{2}$. So, the direction of DW motion induced by ac
current can also be efficiently controlled by applying an external
magnetic field.

\subsection*{IV QUANTUM DEPINNING} \indent

So far, both theoretical and experimental investigations are
mainly focused on the DW's classical behavior induced by a
spin-polarized current, such as the classical depinning,
nucleation \cite{uniform}, and so on. These works show us that the
spin-polarized current can always make the same effects in DW's
behavior with the magnetic field do. As we know that the magnetic
field can induce macroscopic quantum phenomena such as quantum
nucleation and quantum depinning of a DW via macroscopic quantum
tunnelling \cite{tunnel1, tunnel2, nucle}. This motives us to seek
how the spin-polarized current induce the quantum behavior of the
DW. We here concentrate on the quantum depinning of a DW.

Considering the above discussion, the classical dynamic equation of
the DW Euler equation is obtained as
\begin{eqnarray}\label{euler}
m\ddot{q}+Q\dot{A}+=QH_{ext}-\frac{\partial u(x)}{\partial x}.
\end{eqnarray}
In this section, we neglect the damping term and assume that
vector potential $A$ satisfies $A=Kt$, where $K=\frac{m\mu_b
P}{QeM_s}\frac{d j_e}{dt}$, which means the spin-polarized current
rises with time linearly. After wick rotation, the Lagrangian and
Euler equation in the imaginary time are given by
\begin{eqnarray}\label{lagim}
L=-\frac{1}{2}m(i\frac{dq}{d\tau}+b_J)^2+u(x),
\end{eqnarray}
\begin{eqnarray}\label{eulerim}
m\frac{d^2 q}{d\tau^2}-QK-\frac{\partial u}{\partial x}=0.
\end{eqnarray}
Then, according to the standard instanton method ($\tau=it$), the
tunnelling rate is given by
\begin{eqnarray}\label{rate}
\Gamma=C_0\exp(-B_0)  \ \ \ \ \ B_0=\frac{S_{0}}{\hbar}.
\end{eqnarray}
Here $C_0$ is the prefactor and $S_0$ is the action of the
classical solution in imaginary time corresponding to Eq.
(\ref{eulerim}), which is similar with the case of ferromagnetic
applied with an external magnetic field. The classical action
$S_0$ is written as
\begin{eqnarray}\label{action}
&&S_0=\int_{\frac{-\tau_0}{2}}^{\frac{\tau_0}{2}} d\tau
\frac{1}{2m}(m\frac{dq}{d\tau}+QK\tau)^2+u(x)\nonumber
\\
&&=\int_{\frac{-\tau_0}{2}} ^{\frac{\tau_0}{2}} d\tau [(\frac{1}{2}m
(\frac{dq}{d\tau})^2+u(q)-QKq)\nonumber
\\
&&+(QKq+QK\tau\frac{dq}{d\tau})+\frac{1}{2}(\frac{dq}{d\tau})^2],
\end{eqnarray}
where $\tau_0$ is the rising time of the current. The first term
in classical action $S_0$ is the same with of quantum depinning
induced by external magnetic field. The second term is a total
differential which equals to
$QKq(\tau)\tau|^{\tau_0/2}_{-\tau_0/2}$ and the third term equals
to $\frac{1}{3}k\tau_{0}^{3}$. Both the later two terms are
imaginary, so they just add to a phase in the action and has no
effect on the tunnelling rate. Then, we reach our conclusion that
a polarized current also can induce a DW tunnelling from a pinning
potential.

We still consider harmonic potential in Eq. (\ref{clpin}). with
this potential and according to Eq. (\ref{action}), we obtain the
classical action of the DW as
\begin{eqnarray}\label{act0}
S_0=\frac{1}{2} \sqrt{m} \omega \xi(\xi-\frac{2\xi
QK}{m\omega^2}),
\end{eqnarray}
and the prefactor can be calculated straightforwardly \cite{pre}
\begin{eqnarray}\label{pre}
C_0=(\frac{\omega}{\pi\hbar})^{\frac{1}{2}}(2 \sinh(\omega
\tau_0))^{\frac{1}{2}}.
\end{eqnarray}
Thus the quantum tunnelling rate can easily be obtained by the Eq.
(\ref{rate}). The above results show that quantum tunnelling and
nucleation of DW can also be induced by spin-polarized. As the
current is more precise than magnetic field in controlling DW, the
above conclusion is worth of further study in future works.

\subsection*{V. Conclusions} \indent
In conclusion we have studied the classical and quantum depinning of
the one-dimensional domain wall (DW) induced by a fast-varying
spin-polarized current. Firstly, we confirm the adiabatic condition
of the spin-torque in fast-varying current case and predict that the
magnetoresistance will appear two peaks with the frequency' variety.
Then, the Lagrangian of the DW based on the solution in \cite{last}.
It is interesting that an effective vector potential $A(t)$ appears
in the Lagrangian due to the interaction between the DW and the
spin-polarized current. This means the DW motion can be understood
as a charged particle interacts with external magnetic field
described by the vector potential. Together with these results, we
can study the depinning of DW and find that there are two critical
values for the current, $(\frac{d J_e}{d t})_{cr}=\frac{eM_s\xi
\omega^2}{\mu_B P}$ and $J_{cr}=\frac{2\alpha
eM_s^2\xi}{\gamma\Delta\mu_B P}$, to determine the classical
depinning. These results can explain the experiment \cite{sub}.
Furthermore, by using an external magnetic field, we propose a
scheme to lower the threshold current for the DW's depinning and
control its motion direction. Finally, we discuss the quantum
behaviors, and reveal that the quantum tunnelling effect of the DW
can also be induced by the spin-polarized current.

\subsection*{ACKNOWLEDGMENTS} \indent
We thank Prof. Wu-Ming Liu, Prof. Ke Xia and Prof. Jingling Chen
for their valuable discussion. This work is supported by NSF of
China under grants No. 10275036, and by NUS academic research
Grant No. WBS: R-144-000-071-305.




\noindent\\  \\

\end{document}